# Polarized cold cloud of thulium atom

V. V. Tsyganok[1,2], V. A. Khlebnikov[1], E. S. Kalganova[1,2,3], E. T. Davletov[1,2,4], D. A. Pershin[1,2], I. S. Cojocaru[1,2], I. A. Luchnikov[1,2,4], V. S. Bushmakin[1,2],
V. N. Sorokin[1,3], A. V. Akimov[1,3,5]

[1]Russian Quantum Center, Business Center "Ural", 100A Novaya Str., Skolkovo, Moscow, 143025, Russia
[2]Moscow Institute of Physics and Technology, Institutskii per. 9, Dolgoprudny, Moscow Region 141701, Russia
[3]PN Lebedev Institute RAS, Leninsky Prospekt 53, Moscow, 119991, Russia
[4]Center for Energy Systems, Skolkovo Institute of Science and Technology, 3 Nobel Street, Skolkovo, Moscow Region 143026, Russia
[5]Texas A&M University, 4242 TAMU, College Station, Texas, 77843, USA,
email: akimov@physics.tamu.edu

## I. ABSTRACT

Minimization of internal degrees of freedom is an important step in the cooling of atomic species to degeneracy temperature. Here, we report on the loading of $6 \cdot 10^5$ thulium atoms optically polarized at maximum possible magnetic quantum number $m_F = -4$ state into dipole trap operating at 532 nm. The purity of polarizations of the atoms was experimentally verified using a Stern-Gerlach-type experiment. Experimental measured polarization of the state is $m_F = -3.91 \pm 0.26$.

## II. INTRODUCTION

Cold atoms with temperatures equal or close to the temperature of degenerate gas [1] have attracted a lot of attention in recent years due to their potential in various applications ranging from frequency standards and metrology [2,3] to quantum simulation of electro-magnetic properties of solid state materials [4–6], turbulence [7] or even the formation of stars [8].

Lanthanides have a special place in the field of quantum simulations due to their unique properties, such as large orbital momentum [9,10] and large magnetic momentum in the ground state [11,12]. Large orbital momentum in the ground state leads to easily tunable interactions between cold atoms via low-field Feshbach resonances [13], while large magnetic momentum leads to relatively strong dipole-dipole interactions [10]. In particular, the thulium atom has an orbital angular momentum of 3 and magnetic moment of 4 Bohr magnetons in the ground state. In addition, it has a relatively simple level structure, thus combining the strength of more developed Er and Dy [9,14] with the advantage of a clearer system.

Cooling of the thulium atom to the Bose-Einstein condensation (BEC) has not been achieved yet. Traditionally, the last step in cooling to BEC temperatures is done via evaporative cooling, either in a magnetic or in an optical dipole trap (ODT). In the case of the thulium atom, inelastic collisions in a magnetic trap were measured and found to be quite substantial [15]; therefore, we selected the optical dipole trap for realization of evaporative cooling of Tm atoms. In order to avoid spin-exchange collisions, atoms in the trap need to be polarized into magnetic sublevel with minimal energy; i.e., magnetic momentum is maximally aligned with the external magnetic field.

Several methods of polarization of an atomic cloud in a dipole trap were suggested previously. Among them are polarization by optical pumping [10,16] and polarization of atomic ensemble in highly detuned magneto-optical trap (MOT) prior to loading into the dipole trap [17]. The latter method has the advantage of absence of heating in the polarization step, as well as simplicity (since no extra laser or microwave source is required).

In this paper, we report on achieving a polarized atomic cloud of thulium atoms in a dipole trap operating at 532 nm using preloading polarization in far detuned MOT and estimate the purity of the polarization achieved.

## III. POLARIZATION OF THE ATOMIC CLOUD

Thulium atoms were first precooled in a magneto-optical trap operating at 530.7 nm transition in thulium atom (see Figure 1 A, B). Atoms were evaporated in an effusion cell, then slowed down with a Zeeman slower and additionally slowed down with molasses operating at a wavelength of 410.6 nm, and finally captured in a magneto-optical trap operating at 530.7 nm (see previous work [18] for more details). In our trap design, we were able to capture up to $3\cdot 10^7$ atoms at a temperature of 13 µK. To realize polarization of thulium atoms cooled this way, we explored the approach suggested at work [17]. The general idea of this method is to use a highly detuned magneto-optical trap (MOT) to create preferential absorption of one circular polarized beam by a cold atomic cloud displaced from 0 of the magnetic field due to gravitational force. In order to implement such a polarization method after initial cooling in our MOT, we turned off the Zeeman and molasses beams and detuned the MOT beam frequency to the desired detuning (varied in the range $0-25\Gamma$ depending on the experiment). The power of the cooling beams was set to 20 µW (beam radius at $e^{-2}$ level $9.1\,\mathrm{mm}$) leading to peak intensity $I = 2P/\pi\sigma^2 = 0.05 I_{sat}$ ( $I_{sat} = \dfrac{h\pi c\Gamma}{3\lambda^3} = 316\,\mu\mathrm{W/cm^2}$ in one beam) and the gradient of the magnetic field was set to 4.64 G/cm (see APPENDIX B). At such a large detuning, atoms at MOT do not experience a noticeable optical cooling force at any point of the MOT volume except for the narrow shell, at which the magnetic field is enough to compensate for the detuning of the cooling beam (see Figure 1C). Feeling gravitational force atoms initially cooled in the center of the MOT will naturally shift down toward the MOT border described above, filling the bottom part of the shell (Figure 1D). Here, atoms will automatically interact preferably with the upward beam, which creates a contraction for the gravitational force but would be far detuned from the frequency of the downward beam and to a high degree from frequency of the horizontal beams. Such asymmetry creates optical pumping by upward beam into $m_F = -4$ component (atom magnetic moment is directed along the external magnetic field).

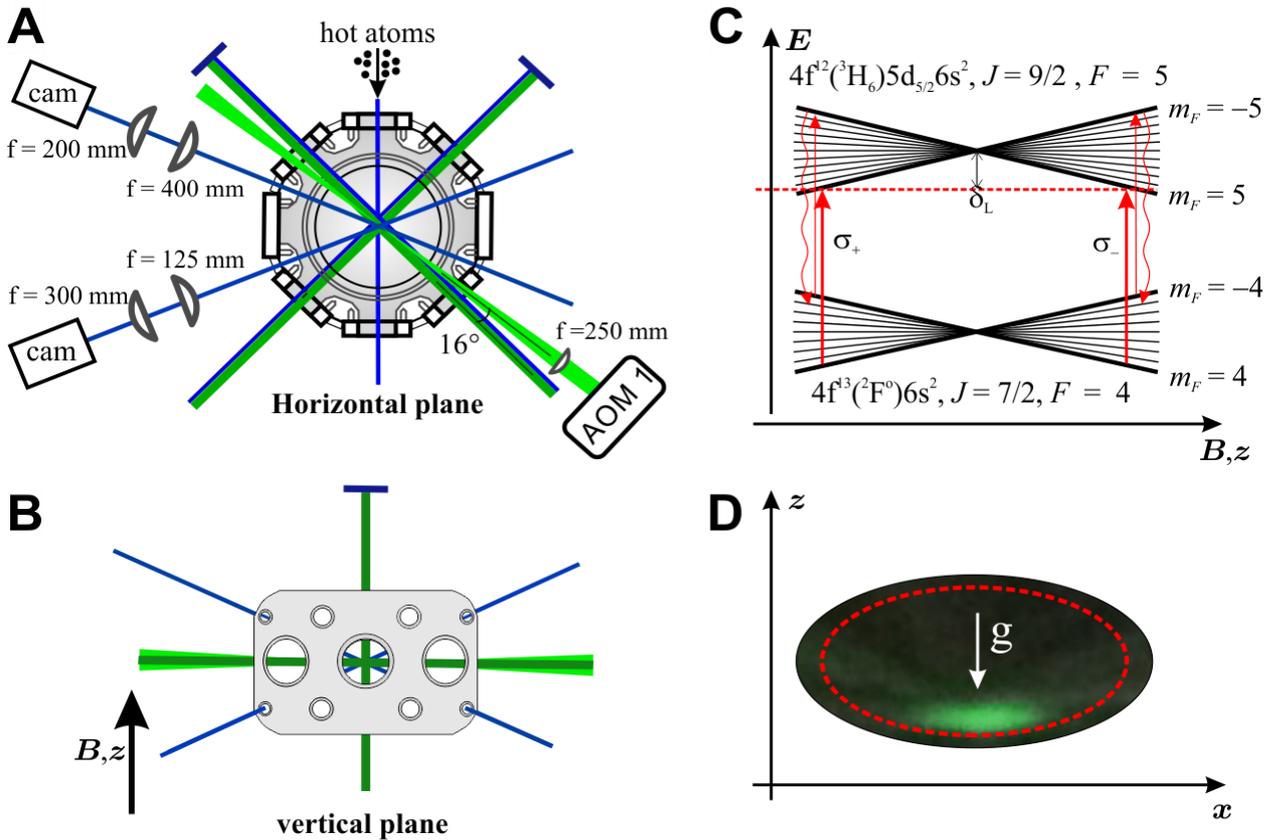

Figure 1 A, B) Experimental setup, top and side view, respectively. C) Structure of the levels of the cooling transition in the magnetic field. Red arrows indicate cooling light at large detuning. D) Photograph of the MOT at $15\Gamma$ detuning. The red dashed line illustrates the effective border of the MOT at which magnetic field compensate for detuning of the cooling light [19].

Laser detuning and intensity are parameters that define the efficiency of polarization. Obviously, higher intensity causes faster optical pumping to the desired state [20], but it also leads to a higher temperature [17] and faster light-assisted collisions [18], which reduce the number of atoms in the trap. Purity of polarization is mostly defined by the magnitude of detuning of the cooling beams. Roughly, higher detuning means higher asymmetry in light forces and thus more polarization. A high degree of polarization prevents atoms from heating/losses caused by dipolar relaxation [21]. However, as higher trap detuning leads to a bigger volume of the trap, transferring into ODT become less efficient. In order to optimize beam parameters, we performed a simulation of MOT in a far-detuned regime (see APPENDIX A). This treats the population at all Zeeman sublevels depending on the beam's saturation parameter and detuning, which can provide useful hints for choosing the right intensity and detuning. It is also consistent with following the Stern-Gerlach measurements of cloud polarization with respect to MOT beam detuning (see below). We ultimately chose parameters $\delta = 15\Gamma$ and $S_0 = I/I_{sat} = 0.05$, at which the number of reloaded atoms is maximal, with minimal MOT temperature. At this parameter, the cloud of thulium in MOT contains 3·10⁷ atoms at temperature 13 μK with almost ideal polarization (see Figure 3A).

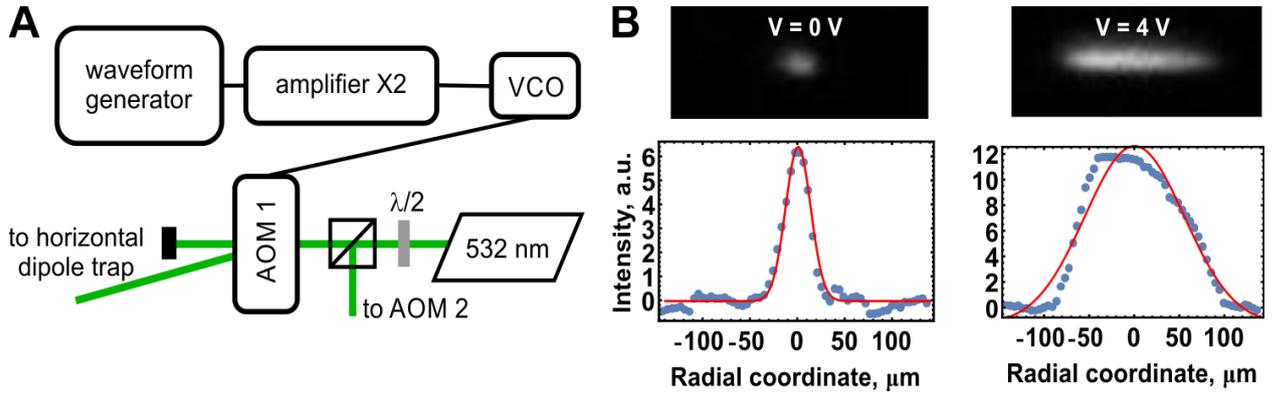

Figure 2 A) Scheme for dipole beam sweeping. B) Measurements of beam waist with no voltage on VCO (left) and with full voltage on VCO (right).

The dipole trap was formed by a laser beam with waists (radius of the beam at $e^{-2}$ level) of $w_x = 23 \mu m$ and $w_y = 15 \mu m$ and a wavelength of 532 nm (see Figure 1A). For our laser source, we used a Coherent Verdi-10 laser. Only 5 W was used in the trapping area. To increase the number of atoms loaded into the dipole trap, the position of the horizontal beam of the dipole trap was swept using an acousto-optic modulator (AOM) as shown in Figure 2A. The sweep was performed by a voltage control oscillator (VCO) DCMO616-5 (mini-circuits feed function generator [Stanford Research Systems DC 345]) with saw waveform. We were able to reach a sweep span of $\delta \nu = \pm 20$ MHz near the central frequency of our AOM of 110 MHz, enabling us to effectively broaden our dipole trap beam by a factor of 4.1 (see Figure 2C). The frequency of the sweep should be high enough to exclude possible heating due to parametric oscillations in the trap [22]. Nevertheless, considerable depolarization was observed at sweep frequencies above 100 kHz, likely due to light-assisted depolarization. The swept dipole trap from the point of view of a stationary atom is intensity modulated light, the splitting of sidebands of which may match with the Zeeman splitting of an atom, thus causing stimulated Raman transitions. Such transitions will decrease atomic polarization and cause additional heating [20]. Therefore, the frequency of 40 kHz—well below 100 kHz but still quite high compared to trap frequencies (which do not exceed 3.5 kHz)—was chosen as the sweeping frequency.

Beam waist measurements were performed with a CCD camera (see Figure 2B). Atoms were loaded into the dipole trap by overlapping the swept dipole trap with the MOT for 300 ms. The wider the beam, the more atoms can be reloaded into a dipole trap due to the simple increase in geometric overlapping with MOT. Sweeping the beam, however, makes the effective potential shallower, a result which was compensated for by utilization of larger trapping beam power. Experimentally, we found that for the maximum aspect ratio of the sweeping beam of 6, the trap has $1.2 \cdot 10^6$ atoms (compared to $1 \cdot 10^5$ atoms without this sweep) at 22 μK and after turning off the sweeping, was turned into $6 \cdot 10^5$ atoms at 16.7 μK.

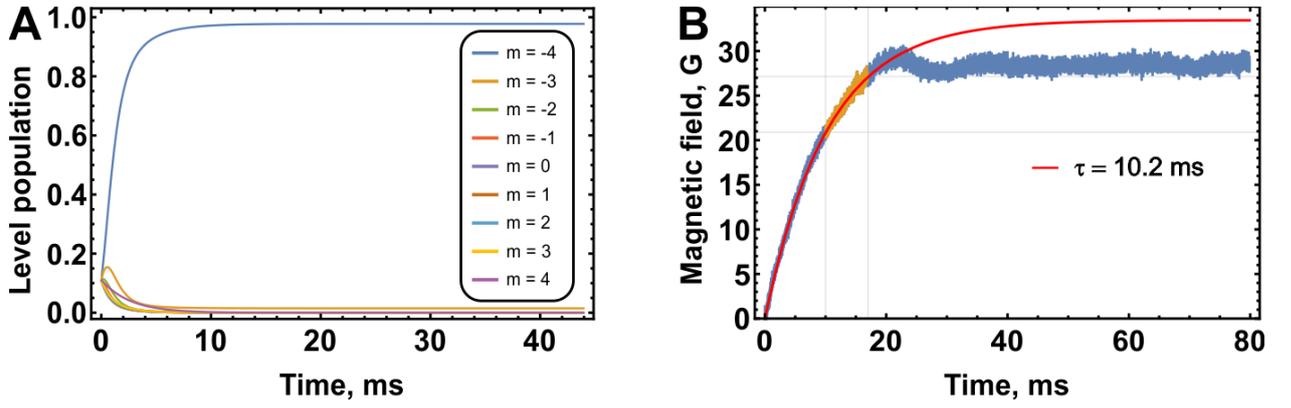

Figure 3 A) Population of the different magnetic sublevels (simulation) versus the time atoms spend in the highly detuned MOT. The following parameters were used: $S_0 = 0.02$, $\Delta = 17\Gamma$. B) Magnetic field gradient rise time measurements. The yellow segment illustrates the time interval at which measurements of the cloud polarization were performed. The red curve illustrates the fit with $\beta(1 - \mathrm{Exp}[-t/\tau])$ dependence in which $\beta$, $\tau$ were the fitting parameters.

## IV. STERN-GERLACH-TYPE EXPERIMENT

To measure the polarization of the atomic cloud, one can choose among several techniques, for example, the Stern-Gerlach-type experiment [23] or absorptive spin-sensitive in-situ imaging [24]. Since MOT naturally uses rather strong magnetic field gradients, the Stern-Gerlach-type experiment was chosen for our measurements. The main idea of this technique is that atoms with different magnetic quantum numbers have different acceleration in the gradient of magnetic field. Therefore, motion and expansion of an atomic cloud in the presence of such a gradient witness the degree of polarization of the cloud. In limiting the case in a sufficiently cold, high magnetic field, cloud should separate into independent clouds corresponding to a specific magnetic quantum number. As the temperature of atoms in the ODT was not low enough to resolve the different magnetic states, we explored the dynamics of cloud waist and center of mass to compare with theoretical expectations. An imaging technique absorption was used.

The experiment was realized in the following way. After reloading atoms from MOT to ODT, a constant "storage" magnetic field was turned on to suppress depolarization collisions [25]. The sweeping amplitude was reduced to 0 with a linear ramp during 50 ms, accompanied by a corresponding reduction of the beam power so that the dipole trap was kept constant during the whole ramp. Then, in part of the experiments, a gradient magnetic field of 29 G/cm formed by the MOT gradient coils, was turned on. The nominal value of the gradient was measured by the position shift of the magnetic trap [26] with the calibrated DC magnetic field (see APPENDIX B). The maximum field gradient is reached only 25 ms after the voltage change at the power source of the coils, which is comparable to the time required for atoms to leave the camera field of view. Therefore, the rise profile of the gradient was measured by Hall probe (Honeywell SS495A) as shown at Figure 3B and calibrated using the previously measured field gradient. Then, the profile was used in the analysis of the cloud expansion (see APPENDIX B). In these experiments, the constant storage magnetic field was maintained with the gradient on, thus causing 0 of the gradient to show well above the cloud and avoiding any possibility of spin-flips at 0 of the field gradient. Then, 10 ms after the gradient was turned on, the horizontal beam was blocked and the free-fall of the atomic cloud in the gradient of magnetic field was observed. The results are shown in Figure 4. Clearly, the center of mass of the atomic cloud motion strongly deviated from the 0 gradient dynamics shown in Figure 4(B,D), but the width of the cloud stayed the same as shown in Figure 4(A,C) in both cases.

The polarization degree of the atomic cloud achieved in the experiment could be estimated from the center of mass acceleration. Assuming single polarization, the change in the center of mass acceleration $a$ in vertical direction $z$ could be described as:

$$a(t) = g + \frac{1}{M_{Tm}} \frac{d\vec{\mu}\vec{B}(t)}{dz} = g + m_F \frac{\mu_B}{M_{Tm}} \frac{dB(t)}{dz} \tag{1}$$

Here, $g$ is acceleration due to gravity, $\vec{\mu}$ is atomic cloud magnetization, $\vec{B}$ is the vector of magnetics field in ODT region, $m_F$ is magnetic quantum number, $\mu_B$ is the Bohr magneton, and $t$ is the time. Since the gradient of the magnetic field was measured independently, the position of the center of mass $z(t)$ is described by:

$$z(t) = \sum_{m_F=-4}^{m_F=4} n(m_F) m_F \int_{t_0}^{t}\left(\int_{t_0}^{t} \Delta a(t')dt'\right)dt' = \langle m_F \rangle \int_{t_0}^{t}\left(\int_{t_0}^{t} \Delta a(t')dt'\right)dt',$$

$$\Delta a(t) = \frac{\mu_B}{M_{Tm}} \frac{dB(t)}{dz} \tag{2}$$

$$\langle m_F \rangle = \sum_{m_F=-4}^{m_F=4} n(m_F) m_F$$

Here $n(m_F)$ is the fraction of atoms with specific $m_F$ and $\langle m_F \rangle$ is the average polarization of the cloud. Since the gradient was not constant during our expansion experiment (see Figure 3B), the magnetic field gradient of the following model was used:

$$\frac{dB}{dz} = \beta\left(1 - \text{Exp}[-t/\tau]\right) \tag{3}$$

Here, $\beta, \tau$ were the fitting parameters. While the model clearly fails in the final gradient value, it adequately describes the field behavior at the time interval of interest. Thus, average polarization of the atomic cloud could be estimated using the cloud expansion, as shown in Figure 4B. The widths have similar information on the degree of polarization. Thus, Figure 4A shows almost no difference in cloud width, both with and without the gradient. Nevertheless, the inset evidence reveals the small presence of the $m_F = -3$ component.

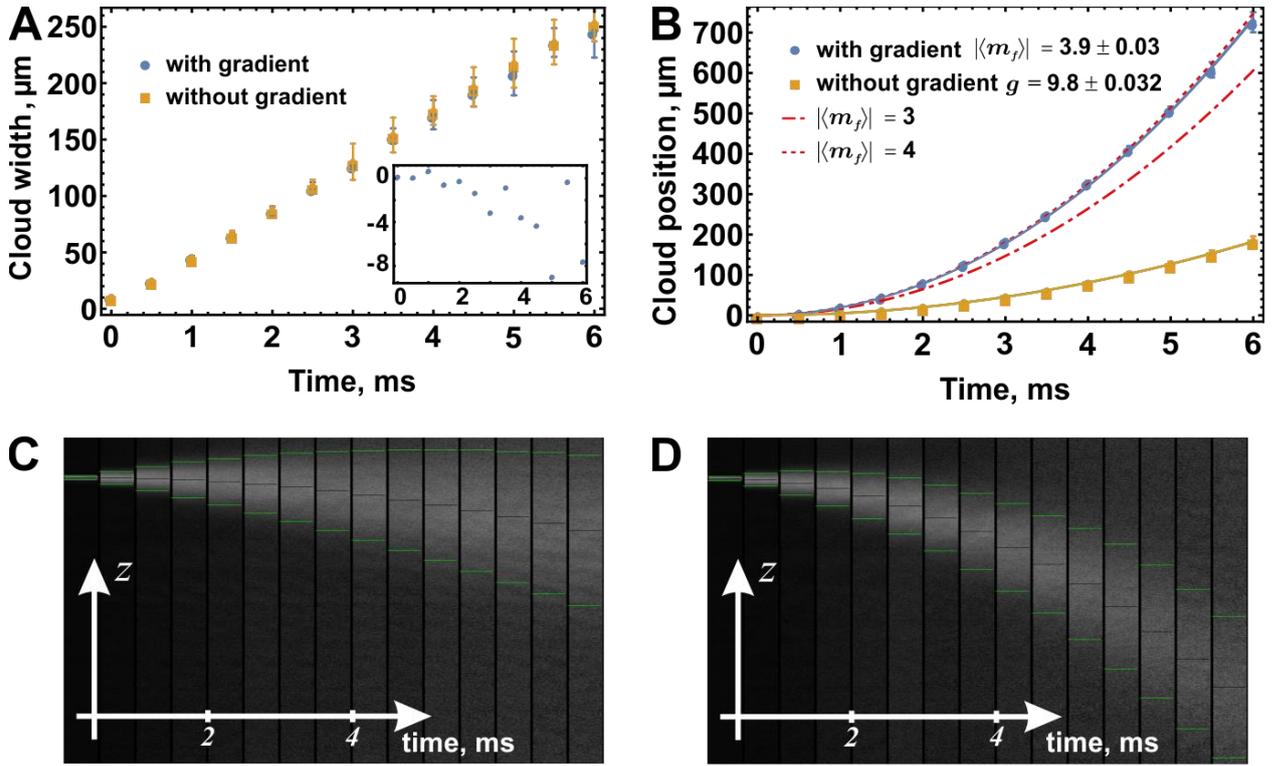

Figure 4 A) Width of the atomic cloud after release from the polarized ODT, with and without the gradient of magnetic field. The inset illustrates the difference between cloud width with and without the gradient. B) Displacement of the atomic cloud after release from the polarized ODT with and without the gradient of the magnetic field. The dashed red line illustrates the calculation of the cloud fully polarized into $m_F = -4$ state; the dot-dashed red line -- $m_F = -3$ and solid line, the fit of the experimental data with $\langle m_F \rangle = -3.9$. C) Photograph of the atomic cloud after polarized ODT is turned off in the absence of the field gradient. The vertical axis is compressed 3 times. D) The same as C, with the gradient. The black horizontal line illustrates the center of mass and green, e$^{-2}$ level.

More insight into the degree of cloud polarization could be gained from measuring the dynamics of the non-polarized atomic cloud. The non-polarized cloud could be achieved by loading ODT from MOT with small detuning of $0.2\Gamma$ and cooling beam power 23 mW. At this configuration, MOT is practically collected at 0 of the magnetic field and therefore does not have noticeable polarization. In this configuration, in the absence of the magnetic field gradient, the atomic cloud expands uniformly, with a slight distortion due to gravitation, as noted in Figure 5(A,B). The addition of a high magnetic field gradient considerably accelerates the expansion rate, but the center of mass of the cloud keeps moving with the same acceleration. This measurement allows verification of the calibration of the field gradient $dB(t)/dz$. Indeed, expression (1) allows one to find the value of the gradient by knowing the acceleration of atoms with each specific $m_F$. This could be done by fitting the expansion of the unpolarized cloud with the sum of 9 equidistant and equal in amplitude (see APPENDIX C) Gaussian profiles (Figure 5C). Using this method, we found that parameter $\beta$ in (3) is 35 G/cm (see Figure 5D) compared to the fit by a measured profile 33.5 G/cm. This discrepancy is within our error bar and is likely due to errors in the linear dimensions calibration and set main uncertainty of the measurement (see APPENDIX C).

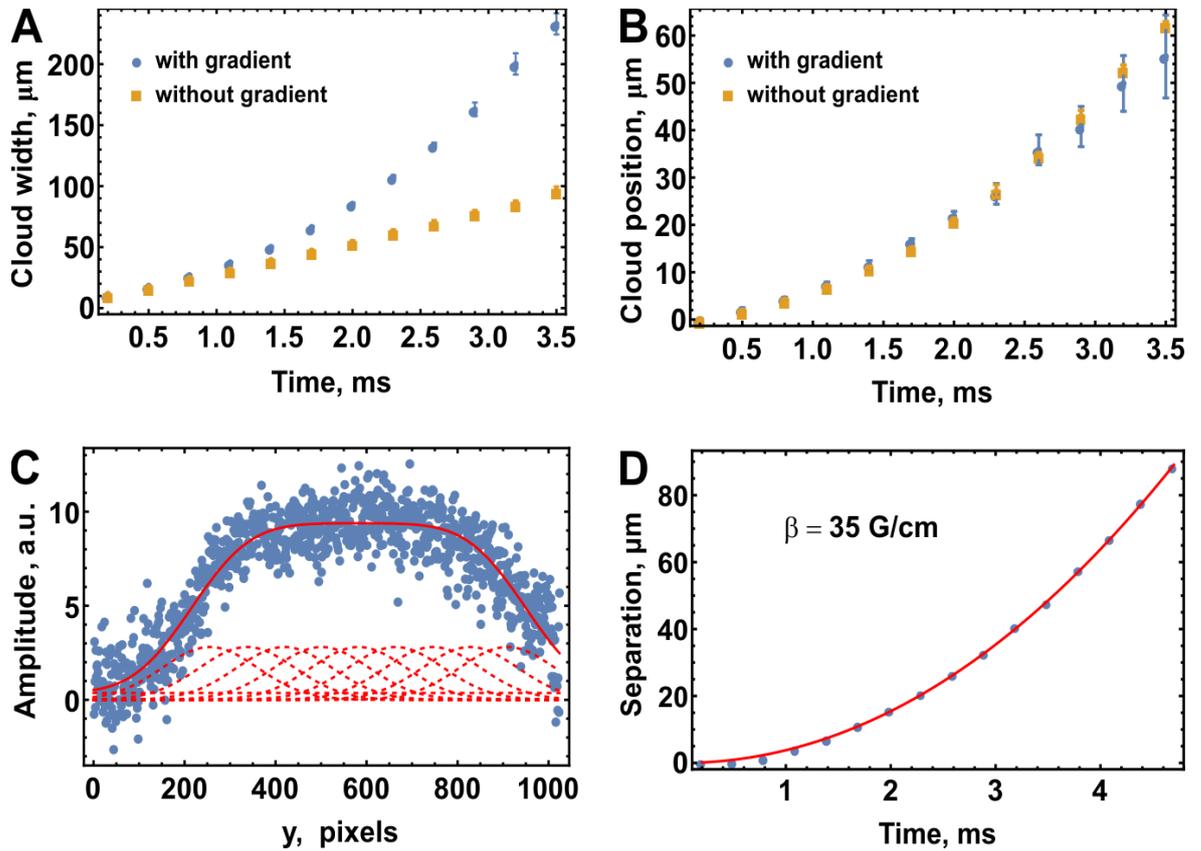

Figure 5 A) Width of the atomic cloud after release from the unpolarized ODT with and without the gradient of the magnetic field. B) Displacement of the atomic cloud after release from the polarized ODT, with and without the gradient of the magnetic field. C) Fit of the unpolarized cloud expansion with 9 Gaussian curves. D) Distance between the atomic clouds and neighbor $m_F$.

Optimization of cloud polarization requires careful selection of the storage magnetic field and parameters of the MOT. Figure 6A illustrates the dependence of ODT polarization on the magnetic field. Small deviations from the purely exponential rise seen around 4 and 5 Gauss are likely due to depolarization collisions. At the same time, light assisted depolarization processes seem to not play a significant role, since no dependence of cloud polarization on the dipole trap laser power was found.

Next, MOT cloud polarization needs to be optimized. Direct measurement of MOT polarization is difficult, since the MOT cloud in high detuning is initially big and continues undergoing thermal expansion once released. Therefore, we performed measurements in ODT, but varied the parameters of the MOT from which the ODT was loaded. Figure 6B shows the experimentally observed dependence on the MOT detuning, which is slightly less steep than the one that was calculated. Experimentally, we found that optimal detuning, providing maximum polarization without considerable loss of number of atoms is $\delta = 15\Gamma, \Gamma = 0.38$ MHz.

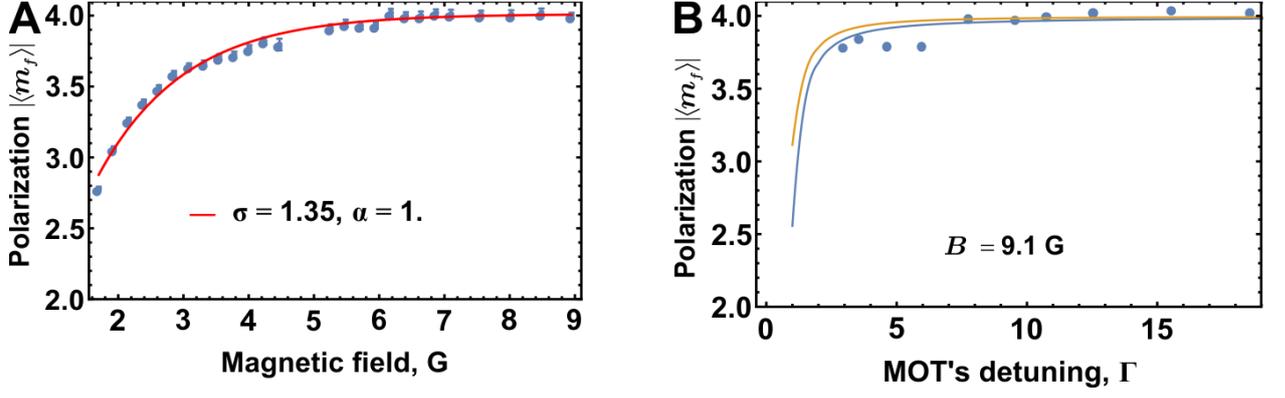

**Figure 6** A) Dependence of ODT polarization on the magnetic field. The red solid line corresponds to fit with $A(1 - \text{Exp}[-(t/\sigma[\tau])^\wedge\alpha])$ function. B) Dependence of the MOT polarization on detuning of the cooling lasers from atomic resonance. The range solid line represents the simulation done for our MOT configuration with $S_0 = 0.02$, and blue shows the correction due to different Clebsch-Gordan coefficients.

## V. CONCLUSION

The polarized cloud of thulium atom in an optical dipole trap operating at 532 nm was demonstrated. The degree of polarization was measured using the Stern-Gerlach-type experiment and optimized experimentally. The maximum achieved polarization was found to be $m_F = -3.91 \pm 0.26$ of $-4$ for $6 \cdot 10^5$ thulium atoms at 16.7 μK in the dipole trap. This step is essential in achieving BEC of the thulium atom.

## VI. AKNOWLEDGEMENTS

This research was supported by Russian Science Foundation grant #18-12-00266.

## APPENDIX A: CALCULATION OF THE ATOMIC POLARISATION IN DETUNED MOT

Thulium atom has a full momentum quantum number of 4 in the ground state and 5 in the excited state, leading to a total of 20 magnetic states. Assuming that the dynamics start from the uniform distribution in the ground state, one can write the initial $20 \times 20$ density matrix $\rho_{in}$ as:

$$\rho_{in} = \frac{1}{9}\begin{pmatrix} O & O \\ O & I \end{pmatrix} \qquad (4)$$

Here, $O$ is zero matrix and $I$ is a unitary $9 \times 9$ matrix. The evolution of the thulium atom once detuning is changed in a stepwise manner as described by the Lindblad equation in rotation wave approximation:

$$\frac{d\rho}{dt} = -\frac{i}{\hbar}[H_0,\rho] - \frac{i}{\hbar}[H_l,\rho] +$$
$$+ \Gamma \sum_k \sum_m \langle 4, m-m_k; 1, m_k | 5, m, 4, 1 \rangle^2 \Big( |m-m_k, g\rangle\langle m, e| \rho |m, e\rangle\langle m-m_k, g| - \qquad (5)$$
$$- \frac{1}{2}\{|m,e\rangle\langle m,e|, \rho\} \Big).$$

Here, $\Gamma$ is spontaneous decay rate, $H_l$ is interaction Hamiltonian, and $H_0$ is atomic Hamiltonian:

$$H_0 = \sum_{i \in g} \mu_B g B m_i |g_i\rangle\langle g_i| + \sum_{i \in e} (\mu_B g B m_i + \hbar\delta) |e_i\rangle\langle e_i|, \qquad (6)$$

with $\mu_B$ — Bohr magneton, $g$ — g-factor, $m_i$ — magnetic quantum number, $g_i$ — ground state levels, $e_i$ — excited state levels, and $B$ — magnetic field, $\delta$ — laser detuning from atomic transition. The interaction Hamiltonian $H_l$ is:

$$H_l = \sum_k \hbar\Omega_k \sum_m \langle 4, m; 1, m_k | 5, m+m_k, 4, 1\rangle |m+m_k, e\rangle\langle m, g| + h.c.. \qquad (7)$$

Here, $\Omega_k$ is Rabi frequency, $m$ — magnetic quantum number, and $k$ — photon state index. The Rabi frequency is calculated using the known polarizations of light and directions of cooling beams.

The solution of equation (5) gives us insight into the level of polarization of the thulium atom and time required in the experiment for achieving the required level of polarization in far detuned MOT. To calculate the average magnetic quantum number, we computed the trace of the product of the density matrix and operator of spin projection $S_z$:

$$S_z = \begin{pmatrix} S_z^{F=5} & O \\ O & S_z^{F=4} \end{pmatrix}$$

$$S_z^{F=N} = \begin{pmatrix} N & 0 & \ldots & 0 & \ldots & 0 & 0 \\ 0 & N-1 & \ldots & 0 & \ldots & 0 & 0 \\ \vdots & \vdots & \ddots & \vdots & \iddots & \vdots & \vdots \\ 0 & 0 & 0 & 0 & 0 & 0 & 0 \\ \vdots & \vdots & \iddots & \vdots & \ddots & \vdots & \vdots \\ 0 & 0 & \ldots & 0 & \ldots & -N+1 & 0 \\ 0 & 0 & \ldots & 0 & \ldots & 0 & -N \end{pmatrix}. \qquad (8)$$

To optimize parameters of the MOT, we also included in consideration temperature of the MOT $T$. According to ordinary Doppler theory [17], the final temperature of MOT is given as

$$T = \frac{D_p}{k_B |\alpha|} = \frac{\hbar\Gamma\sqrt{s_0}}{4 k_B} \frac{R}{\sqrt{R - s_0'/s_0 - 1/s_0}} \qquad (9)$$

Where $D_p$ is the diffusion coefficient, $R = \frac{\hbar k \Gamma}{2mg}$, $s_0'$ is the saturation parameter that takes care of the saturation effects induced by beams in the other direction. However, this regime $s_0' = s_0$ is the saturation parameter of just one beam.

## APPENDIX B: THE FIELD GRADIENT

The DC magnetic field was calibrated using a magnetic field sensor (Lutron Electronic Enterprise Co. Model GU-3001) as follows, First, the coils creating the DC magnetic field were removed from the setup. One of the coils was set on non-magnetic table and the field was measured on the axis of magnetic field with the magnetometer. The error of this measurement is determined by the accuracy of the distance measurements and is about 6%.

Because of the self-induction in magnetic field gradient coils and the limited power of the power supply, the turn-on time for the gradient was about 25 ms (Figure 2B). One could try just to wait this 25 ms to let the gradient rise, but since quadrupole MOT coils were used, the gradient is horizontal as well as vertical in direction, including the direction along the ODT beam. Since confinement along the beam in ODT is weak, 25 ms after the gradient was turned on, more than half of the atoms in ODT flew from the trap in a horizontal direction. Thus, the number of atoms decreased and the cloud size grew and was not suitable for imaging. To reduce this effect, we chose a waiting time of only 10 ms from turning on the gradient to releasing the atoms. This unavoidably led to the gradient changing in time during the expansion experiment. This was done by approximating the gradient with (3), and then, from (1) and (2) one could get for the center of mass position:

$$z(t) = z_0 - \frac{g \cdot t^2}{2} - \beta \cdot \frac{\mu_b \cdot \langle m_F \rangle}{m_{th}} \left[ \frac{t^2}{2} - \tau^2 \cdot Exp\left(-\frac{t+t_0}{\tau}\right) - \tau \cdot t \cdot Exp\left(-\frac{t_0}{\tau}\right) + \tau^2 \cdot Exp\left(-\frac{t_0}{\tau}\right) \right] \quad (10)$$

Where $t_0$ is offset between switching on the gradient and switching off the ODT, $\mu_b$ - Bohr magneton and $\tau = 10.2 ms$ is gradient rising time, the value of which was taken from the fit illustrated in Figure 3.

## APPENDIX C: MEASUREMENT UNCERTANTY

There are two sources of uncertainty during the imaging process. The main one relates to the different Clebsch-Gordan (CG) coefficients for different transitions and the second is due to the finite detuning of the probe beam. Since an atom emits more photons if the probe beam frequency is closer to atomic resonance and less if it is detuned, atoms experiencing a different Zeeman shift from the exact transition will have different levels of brightness for a given detuning of the probe light. In the case of a spin-polarized atomic cloud, the probe beam is in resonance with sigma-minus-polarized light and the

"storage" magnetic field produces a detuning of 1.35Γ for π polarization and 2.7Γ for sigma-plus-polarized light. This difference in detuning is sufficient to suppress absorption, and corresponding fluorescence by one order of magnitude for π polarization and almost two orders for sigma-plus-polarized. Thus, only sigma-minus transitions contribute to the signal. For sigma-minus-polarized light, these coefficients are significantly different for different ground state components; for example, for transition $m_F = -4 \to m_F = -5$ transition, the coefficient is 0.82, for $m_F = -3 \to m_F = -4$, it is 0,65, for $m_F = -2 \to m_F = -3$ is 0.51, and so on. This leads to offset $\delta m_F$ toward bigger polarization, which is summarized in the table below:

| $\langle m_F \rangle$ | -3.95 | -3.90 | -3.85 | -3.81 | -3.71 | -3.57 | -3.23 |
|---|---|---|---|---|---|---|---|
| $\delta m_F$ | 0.016 | 0.031 | 0.044 | 0.06 | 0.078 | 0.1 | 0.18 |
| $\delta m_F / \langle m_F \rangle$, % | 0.41 | 0.79 | 1.1 | 1.6 | 2.1 | 2.8 | 5.6 |

While this is a rather large shift, it end ups being small compared to other uncertainties. Also, due to significant optical pumping during the measurement time (300 ms), by the end of the measurement most of the atoms were pumped into $m_F = -4$ state, thus further reducing shift.

In case of expansion of a non-polarized atomic cloud, Clebsch-Gordan coefficients may cause varying degree of visibility of the different components due to a similar effect. But in this case, the experiment was done in a 0 magnetic field and all transitions were in resonance with the probe beam. In this situation, during exposure time, all atoms will optically pump into state (defined by light polarization) and produce almost the same signal. The time needed for pumping defines the error in this case, for in small detuning it does not exceed 3% between states with maximum different absorption. Therefore, taking 9 gauss with the same amplitude is a reasonable approximation for what is shown in Figure 5C.

The gradient itself was estimated using a magnetometer in the coils and a shift of the purely magnetic trap with the calibrated magnetic field. The main error in the field gradient is geometrical factors. The magnetic field measurement had uncertainty of about 6% (due to the precision of the magnetometer placement). The coordinates of the camera calibration first were estimated using the known magnification, and the final calibration was made using the center of mass motion in the gravitational field, which led to an approximately 3% correction in geometrical factors. Thus, we estimate the uncertainty of the field gradient to be within 6.7%. The difference between the non-polarized cloud expansion experiment and direct measurement of the gradient is 4.5%, as described in the main text, and is smaller than the estimated uncertainty.

Summing up all sources of uncertainty, the final uncertainty for the measured average polarization is no more than 6.7%. The error is slightly asymmetric toward the smaller $\langle m_F \rangle$, but given the overall value, asymmetry is negligible.